\documentstyle[epsfig]{elsart}
\begin{document}

\newcommand{\re}{\mathop{\mathrm{Re}}}
\newcommand{\im}{\mathop{\mathrm{Im}}}
\newcommand{\I}{\mathop{\mathrm{i}}}
\newcommand{\D}{\mathop{\mathrm{d}}}
\newcommand{\E}{\mathop{\mathrm{e}}}

\def\lambar{\lambda \hspace*{-5pt}{\rule [5pt]{4pt}{0.3pt}} \hspace*{1pt}}

{\Large  DESY 13-147}

{\Large  August 2013}

\bigskip

\bigskip

\bigskip

\begin{frontmatter}

\journal{Phys. Rev. ST-AB}

\date{}

\title{Obtaining high degree of circular polarization at X-ray FELs via a reverse undulator taper}

\author{E.A.~Schneidmiller}
\author{and M.V.~Yurkov}

\address{Deutsches Elektronen-Synchrotron (DESY),
Notkestrasse 85, D-22607 Hamburg, Germany}

\begin{abstract}
Baseline design of a typical X-ray FEL undulator assumes a planar configuration which results in a linear polarization
of the FEL radiation. However, many experiments at X-ray FEL user facilities would profit from using a circularly polarized radiation.
As a cheap upgrade one can consider an installation of a short helical (or cross-planar) afterburner, but then one should have an efficient
method to suppress powerful linearly polarized background from the main undulator. In this paper we propose a new method for
such a suppression: an application of the reverse taper in the main undulator.
We discover that in a certain range of the taper strength, the density modulation (bunching) at saturation is practically
the same as in the case of non-tapered undulator while the power of linearly polarized radiation is suppressed by orders of magnitude. Then strongly
modulated electron beam radiates at full power in the afterburner.
Considering SASE3 undulator
of the European XFEL as a practical example, we demonstrate that soft X-ray radiation pulses with peak power in excess of 100 GW and
an ultimately high degree of
circular polarization can be produced. The proposed method is rather universal, i.e. it can be used at SASE FELs and
seeded (self-seeded) FELs, with any wavelength of interest,
in a wide range of electron beam parameters, and with any repetition rate. It can be used at different X-ray FEL facilities, in particular
at LCLS after installation of the helical afterburner in the near future.
\end{abstract}

\end{frontmatter}

\baselineskip 20pt

\clearpage

\section{Introduction}

Successful operation of X-ray free electron lasers (FELs) \cite{flash,lcls,sacla}, based on
self-amplified spontaneous emission (SASE) principle \cite{ks-sase},
opens up new horizons for photon science. One of the important requirements of FEL users in the near future will be polarization
control of X-ray radiation. Baseline design of a typical X-ray FEL undulator assumes a planar configuration
which results in a linear polarization
of the FEL radiation. However, many experiments at X-ray FEL user facilities would profit from using a circularly polarized radiation.
There are different ideas \cite{k-j,ding-huang,bend-polar-1,bend-polar-2,second-harm,fourth-harm,tanaka,geloni-pol,slits}
for possible upgrades of the existing (or planned) planar undulator beamlines.

As a cheap upgrade one can consider an installation of a short helical afterburner. In particular, an electromagnetic
helical afterburner will be installed behind the soft X-ray planar undulator SASE3 of
the European XFEL. However, to obtain high degree of circular polarization one needs to suppress (or separate) powerful linearly
polarized radiation from the main undulator. Different options for such a suppression (separation) are considered:
using achromatic bend between planar undulator and helical afterburner \cite{bend-polar-1,bend-polar-2};
tuning resonance frequency of the afterburner to
the second harmonic of the planar undulator \cite{second-harm};
separating source positions and using slits for spatial filtering \cite{slits}.

In this paper we propose a new method for suppression of the linearly polarized background from the main undulator: application
of the reverse undulator taper. In particular, in the case of  SASE3 undulator
of the European XFEL, we demonstrate that soft X-ray radiation pulses with peak power in excess of 100 GW and
an ultimately high degree of circular polarization can be produced. As for a comparison with the other methods, our suppression method
is free, easy to implement, and the most universal: it can be used at SASE FELs and
seeded (self-seeded) FELs, with any wavelength of interest,
in a wide range of electron beam parameters, and with any repetition rate. It can be applied at different X-ray FEL facilities, in particular
at LCLS after installation of the helical afterburner in the near future.

\section{Method description}

\begin{figure}[tb]

\includegraphics[width=.9\textwidth]{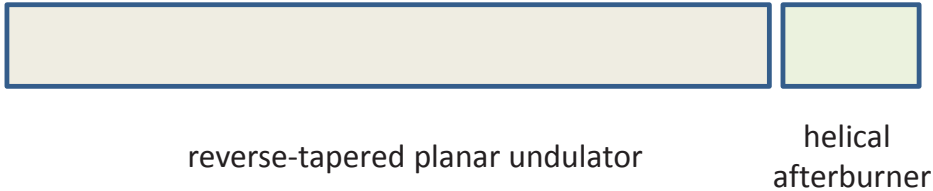}

\caption{\small Conceptual scheme for obtaining circular polarization at X-ray FELs.}

\label{scheme}
\end{figure}

In a short-wavelength SASE FEL the undulator tapering is used for two purposes: to compensate an electron beam energy loss
in the undulator due to the wakefields and spontaneous undulator radiation;
and to increase FEL power (post-saturation taper).
In both cases the undulator parameter K decreases along the undulator length.
The essence of our method is that we use the opposite way of
tapering: parameter K increases what is usually called reverse (or negative) taper.  We discover that in some range of the taper strength,
the bunching factor at saturation is practically the same as in the reference case of the non-tapered undulator, the saturation length
increases slightly while the saturation power is
suppressed by orders of magnitude. Therefore, our scheme is conceptually very simple (see Fig.~\ref{scheme}):
in a tapered main (planar) undulator
the saturation is achieved with a strong microbunching and a suppressed radiation power, then the modulated beam radiates at full
power in a helical afterburner, tuned to the resonance.

Note that reverse undulator taper was considered in the past to increase saturation efficiency of FEL oscillators \cite{oscill-neg},
and to produce attosecond pulses in X-ray FELs \cite{atto-chirp}. In this paper we discover a new useful feature of the reverse taper:
a possibility to generate a strongly modulated electron beam at a pretty much reduced level of the radiation power.

To be specific, in this paper we will concentrate on the case of a helical afterburner and
use the following formula for the degree of circular polarization:

\begin{equation}
D_{\mathrm{cir}} \simeq 1 - \frac{P_{\mathrm{lin}}}{2P_{\mathrm{cir}}} - F_{\mathrm{A}},
\label{degree}
\end{equation}

\noindent where $P_{\mathrm{lin}}$ is the power of the linearly polarized radiation from the main undulator,
$P_{\mathrm{cir}}$ is the power of the circularly polarized radiation from the helical afterburner. Factor of two in the denominator is easy
to understand since the linearly polarized wave can be decomposed into left and right circularly polarized waves, and we consider
the case when $P_{\mathrm{lin}} \ll P_{\mathrm{cir}}$. Except for a contamination due to linearly polarized background from the main undulator,
a decrease of $D_{\mathrm{cir}}$ can be caused by field imperfections of the helical afterburner as well as by other sources of
radiation of the modulated beam (edge radiation, coherent synchrotron radiation etc.) having different polarization properties.
We describe all these possible contributions with a separate term $F_{\mathrm{A}}$. Note that even in the case of an ideal undulator,
the term $F_{\mathrm{A}}$ can be of the order of inverse number of periods in the afterburner.
Further discussions on this subject
would go beyond the scope of this paper since our goal is to minimize the term $P_{\mathrm{lin}}/(2P_{\mathrm{cir}})$. We only notice here that
there is not much sense to make it significantly smaller than the term $F_{\mathrm{A}}$. In most cases it means that a
suppression of
the term $P_{\mathrm{lin}}/(2P_{\mathrm{cir}})$ to a few per mil level is sufficient. Then we can state that
a suppression scheme provides an ultimately high degree of circular polarization.

\section{Selected results of the one-dimensional theory}

A detailed theoretical analysis of the considered effect will be published elsewhere \cite{strong-taper}.
Here we present some selected results.

Let us consider the normalized detuning parameter \cite{book}:

\begin{equation}
\hat{C} = \left( k_{\mathrm{w}} -  \frac{\omega (1+K^2)}{2c\gamma^2} \right) \Gamma^{-1} \ .
\label{detuning}
\end{equation}

The following notations are introduced here: $k_{\mathrm{w}}= 2\pi/\lambda_{\mathrm{w}}$ is the undulator wavenumber,
$\omega$ is the frequency of the electromagnetic wave,
$K$ is the
rms undulator parameter, $\gamma$ is relativistic factor, and $\Gamma$ is the gain parameter. The latter can be expressed in terms of
the FEL parameter $\rho$ \cite{bonifacio}: $\Gamma = 4\pi \rho/\lambda_{\mathrm{w}}$.

We start our consideration with the case when a high-gain FEL is coherently seeded at a given frequency $\omega$, and the undulator is
not tapered. The properties of the FEL are then described by the detuning parameter (see, for example,  \cite{book} for more details).
In particular, in high gain linear regime (i.e. when the normalized undulator length $\hat{z} = \Gamma z \gg 1$) the squared
modulus of the bunching factor $|b|^2$ and the normalized FEL power
$\hat{\eta} = P/(\rho P_{\mathrm{beam}})$  are of the same order when an FEL operates
close to the resonance, $|\hat{C}| < 1$. The normalized growth rate (inverse field gain length)
of the FEL instability, $\re \hat{\Lambda} = \re \Lambda/\Gamma$,
is of the order of unity in this regime.
At the same time, the initial problem solution leads to an interesting result for a large negative detuning, $\hat{C} < 0$ and
$|\hat{C}| \gg 1$. In this case the bunching factor  and the normalized FEL power
are connected in the high gain linear regime by a simple relation:

\begin{equation}
|b|^2  \simeq |\hat{C}|^2  \hat{\eta} \ ,
\label{bunching-ss}
\end{equation}

\noindent i.e. the power is strongly suppressed with respect to the squared modulus of the bunching factor. Note that the tendency
approximately holds at the FEL saturation. The normalized growth rate in the considered case gets smaller,
$\re \hat{\Lambda} \simeq |\hat{C}|^{-1/2}$, with the corresponding increase of the saturation length.

Now let us consider a SASE FEL with linearly tapered undulator. The normalized detuning parameter changes as follows:

\begin{equation}
\hat{C} (\hat{z}) = \beta \hat{z} \ ,
\label{det-taper}
\end{equation}

\noindent where

\begin{equation}
\beta =  - \frac{\lambda_{\mathrm{w}}}{4\pi\rho^2} \ \frac{K(0)}{1+K(0)^2} \ \frac{d K}{d z} \ ,
\label{beta}
\end{equation}

\noindent and $K(0)$ is the initial value of the rms undulator parameter. Note that as a reference frequency we always
consider the resonance frequency at the undulator entrance. Of course, in a SASE FEL a finite frequency band is amplified, and its
maximum and width can evolve along the undulator length \cite{stup}.

The theory of a high-gain FEL  with varying undulator parameter has been developed in \cite{stup} in the limit of a small taper
strength\footnote{\footnotesize More strictly, the condition $|\beta| \hat{z} \ll 1$ is necessary.},
$|\beta| \ll 1$. In particular, the authors of \cite{stup} have derived corrections to the FEL growth rate up to the second order.
Unfortunately, we cannot use the results of \cite{stup} for our purpose because, in the case of small corrections, the tendency we would
like to demonstrate (small ratio $\hat{\eta}/|b|^2$) is not seen. For this reason we present here
a result of the theory \cite{strong-taper} that is valid in the case of a large taper strength. For a high-gain linear regime
and a large negative
taper strength, $\beta <0$ and $1 \ll |\beta| \ll \hat{z}$, the relation between the ensemble-averaged squared modulus of
the bunching factor $<|b|^2>$ and the ensemble-averaged normalized FEL power $<\hat{\eta}>$ can be approximated as

\begin{equation}
<|b|^2>  \simeq |\beta|^{2}  \hat{z}^2  <\hat{\eta}> \ .
\label{bunching-eta}
\end{equation}

\noindent This equation looks similar to Eq.~(\ref{bunching-ss}) with the detuning parameter given by (\ref{det-taper}).
One can see that, indeed, for large negative $\beta$ and large $\hat{z}$, the squared bunching factor is much larger than the
normalized FEL power. Both quantities are proportional to $\exp(4\sqrt{\hat{z}/|\beta|})$, i.e. they
evolve along the undulator length with a decreasing growth rate.

\begin{figure}[tb]

\includegraphics[width=.6\textwidth]{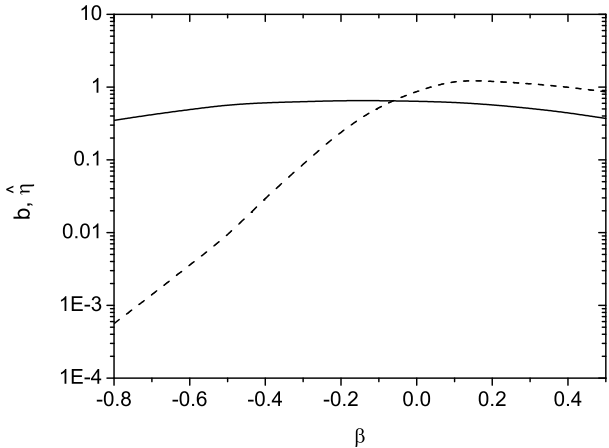}

\caption{\small Ensemble averaged rms bunching factor (solid) and normalized FEL efficiency (dash) at saturation point (position
with maximum bunching factor) versus taper strength parameter. Energy spread parameter is $\hat{\Lambda}_{\mathrm{T}} = 0.2$.}
\label{bunching-1d}
\end{figure}

\begin{figure}[tb]

\includegraphics[width=.6\textwidth]{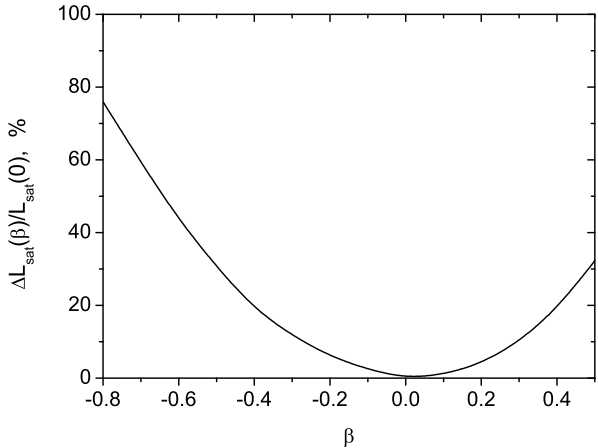}

\caption{\small
Relative increase of the saturation length (defined as a length of the undulator at which maximum bunching is achieved) versus
taper strength parameter. Energy spread parameter is $\hat{\Lambda}_{\mathrm{T}} = 0.2$.
}
\label{sat-length}
\end{figure}

The asymptote of large negative $\beta$ was considered here only for illustration of the power suppression effect. For practical
applications we will restrict ourselves to moderate values of $\beta$ which allow for a significant power suppression at
strong bunching and an acceptable increase of the saturation length. We are interested in the values of bunching factor and FEL power
at saturation, therefore we have to use a numerical simulation code (a linear analysis is not valid at saturation). Below
we will use a simplified notation $b$ instead of $\sqrt{<|b|^2>}$. To make our results of
1-D simulation closer to practical cases, we also introduce an energy spread with a value typical for X-ray FELs. The energy spread parameter
is defined as follows \cite{book}: $\hat{\Lambda}_{\mathrm{T}} = \sigma_{\gamma}/(\gamma \rho)$ with $\sigma_{\gamma}$ being the
energy
spread (in units of the rest energy). In our simulations we use the value $\hat{\Lambda}_{\mathrm{T}} = 0.2$. The results of
simulations
with 1-D version of the code FAST \cite{fast} are presented in Figs.~\ref{bunching-1d} and \ref{sat-length}.

In Fig.~\ref{bunching-1d} we show the bunching factor and normalized FEL efficiency at saturation point which is defined here as
the position where the maximum bunching is reached. One can see that for negative $\beta$ the power quickly decreases (in contrast with
positive $\beta$) although the
bunching factor changes only slightly. From Fig.~\ref{sat-length} one can find how the saturation length depends on the taper
strength. A good range of this parameter for the proposed scheme is $\beta \simeq -0.5 ... -0.3$. Indeed, the bunching factor is still
high in this range, there is only moderate increase of the saturation length, and the power is significantly suppressed.

\section{Three-dimensional simulations for the European XFEL}

The results of the previous Section were obtained in the framework of 1-D model.
We found that the reverse taper method works well in 3-D case, and can even be more efficient than in 1-D case. We illustrate this
with the parameters of the soft X-ray SASE3 undulator of the European XFEL \cite{euro-xfel-tdr}.
Main parameters used in our simulations are
presented in Table~1. The electron beam parameters are taken from the table provided by the European XFEL beam dynamics group
\cite{beam-param}
for the bunch charge of 0.5 nC.
We consider operation of SASE3 in "fresh bunch" mode \cite{fresh} when the energy spread of electron bunches is not
spoiled by the FEL interaction in the upstream SASE1 undulator.
The simulations were performed with 3-D version of the code FAST \cite{fast}.

A gap-tunable permanent-magnet SASE3 undulator consists of 21 undulator modules, each of them is 5 m long.
One can easily control active part of the
undulator by opening the gaps of the modules which are not needed. In our case we use only 11 last modules to adapt an active undulator
length to the
saturation length for the given wavelength (1.5 nm) and electron beam parameters. A long-period electromagnetic helical
afterburner is being developed \cite{binp}
for installation behind SASE3 undulator. The choice of technology is driven by the request of users to quickly change (between
the macropulses, i.e. with the frequency of 5 Hz) the polarization of the output radiation between left and right.

\begin{table}[tb]
\caption{Main parameters used in simulations}
\bigskip

\footnotesize

\begin{tabular}{ l l }
\hline \\
{\bf Electron beam} &  \\
Energy & 14 GeV  \\
Charge & 0.5 nC \\
Peak current	 &  5 kA        \\
Rms normalized slice emittance	 &  0.7 $\mu$m     \\
Rms slice energy spread  &  2.2 MeV   \\
{\bf Planar undulator} &  \\
Period  & 6.8 cm \\
$K_{\mathrm{rms}}$ &  5.7 \\
Beta-function & 15 m \\
Active magnetic length & 55 m \\
Taper $\Delta K_{\mathrm{rms}}/K_{\mathrm{rms}}(0)$ & 2.1 \% \\
{\bf Helical afterburner} &  \\
Period  & 16 cm \\
$K$ &  3.6 \\
Beta-function & 15 m \\
Magnetic length & 10 m \\
{\bf Radiation} &  \\
Wavelength & 1.5 nm \\
Power from planar undulator, $P_{\mathrm{lin}}$  & 0.4 GW \\
Power from helical undulator, $P_{\mathrm{cir}}$  & 155 GW \\
$1 - P_{\mathrm{lin}}/(2P_{\mathrm{cir}})$ & 99.9 \% \\
\hline \\
\end{tabular}

\label{tab:param}
\end{table}

We optimized the taper strength in the main undulator such that the radiation power is sufficiently suppressed, on the one hand, and
the bunching factor is still close to that in the case of untapered undulator, on the other hand. We ended up with
2.1 \% increase of K parameter over the undulator length of 55 m. According to (\ref{beta}), this corresponds to the
one-dimensional\footnote{Note that in the considered case the one-dimensional normalization is not very convenient since the
diffraction effects play a major role.
If 1-D parameter $\rho$ in (\ref{beta}) is substituted by the corresponding 3-D parameter \cite{book},
the taper strength is then -0.4.} normalized taper strength of -0.34.

\begin{figure}[tb]

\includegraphics[width=.6\textwidth]{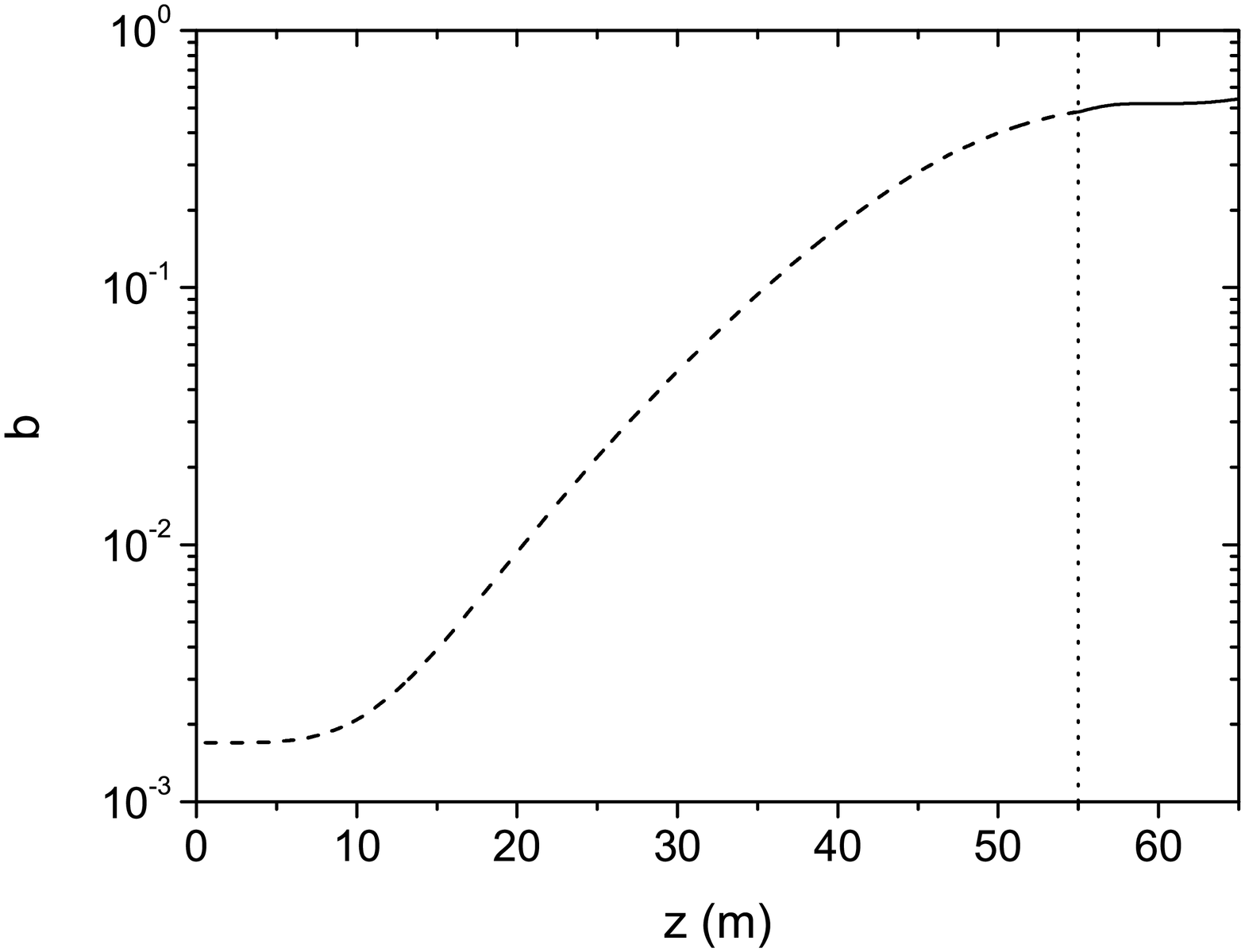}

\caption{\small Evolution of the ensemble averaged rms bunching factor along the planar undulator SASE3 (dash) and the helical
afterburner (solid).}

\label{bunching-sase3}
\end{figure}

\begin{figure}[tb]

\includegraphics[width=.6\textwidth]{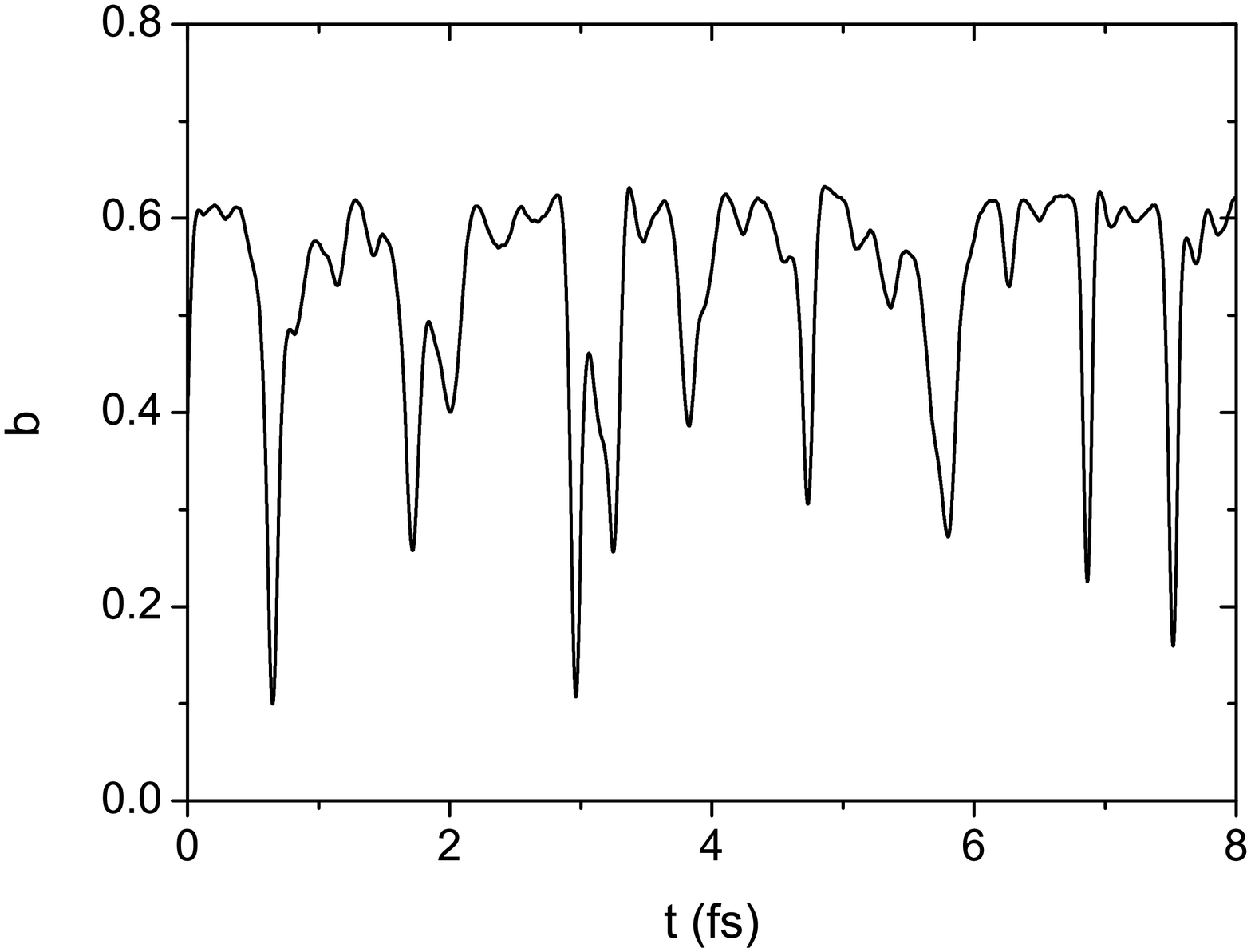}

\caption{\small Modulus of bunching factor versus time at the exit of the planar undulator SASE3
(position 55 m on Fig.~\ref{bunching-sase3}). A central part of the electron bunch is shown.}

\label{bunching-sase3-time}
\end{figure}

Evolution of the bunching factor along the planar undulator and the helical afterburner is
shown in Fig.~\ref{bunching-sase3}, and the time dependence of bunching factor at the exit of the planar undulator - in
Fig.~\ref{bunching-sase3-time}. One can see that the bunching factor reaches a pretty high level and
becomes even larger in the helical afterburner.

\begin{figure}[tb]

\includegraphics[width=.6\textwidth]{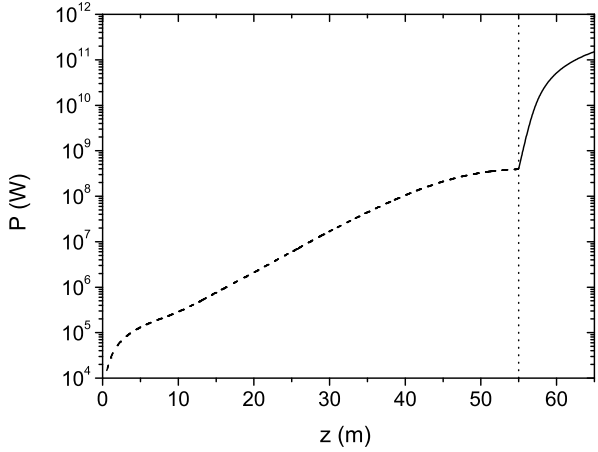}

\caption{\small FEL power versus the length of the planar main undulator SASE3 (dash) and the helical
afterburner (solid).}

\label{power-sase3}
\end{figure}

\begin{figure}[tb]

\includegraphics[width=.6\textwidth]{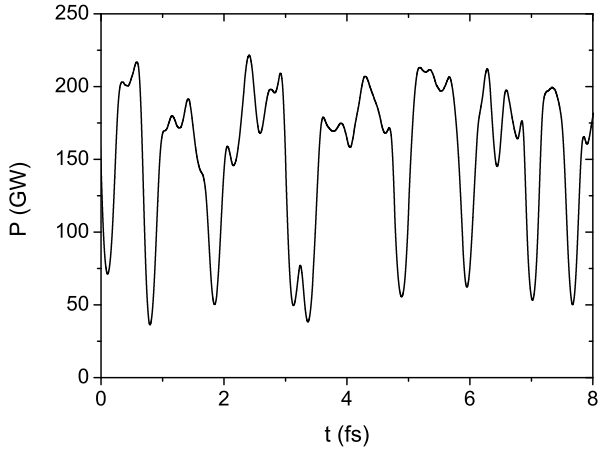}

\caption{\small Peak power of circularly polarized radiation at the exit of the afterburner (position 65 m on Fig.~\ref{power-sase3}).
A central part of the X-ray pulse is shown.}
\label{power-sase3-time}
\end{figure}

Radiation power as a function of position in the planar main undulator and in the helical afterburner is shown in
Fig.~\ref{power-sase3}. One can see that, indeed, linearly polarized radiation from the main undulator is strongly suppressed (it
is about 0.4 GW),
and the powerful circularly polarized radiation quickly builds up in the afterburner. This happens because the bunching is strongly
detuned from the resonance with the last part of the planar undulator, but the $K$ value of the afterburner is optimized
in such a way that it is close to the resonance, and maximum power is achieved at the end of the afterburner. A part of the
radiation pulse is shown in Fig.~\ref{power-sase3-time} for illustration; ensemble averaged peak power reaches 155 GW.
Now we can calculate the degree of circular polarization (not considering the term $F_{\mathrm{A}}$) due to contamination from
the planar undulator: $1 - P_{\mathrm{lin}}/(2P_{\mathrm{cir}}) \simeq 0.999$.
Note that a further suppression of the linearly polarized background and improvement of the quantity
$1 - P_{\mathrm{lin}}/(2P_{\mathrm{cir}})$ is easily possible by going to a stronger taper at the price of
a mild reduction of bunching factor (and, consequently, the power of circularly polarized radiation). However, this would probably
make no sense because the degree of circular polarization would be mainly defined by the term $F_{\mathrm{A}}$, see the discussion
above.

Parameters of the helically polarized radiation are shown in Table 1. The pulse duration and the pulse energy are
defined by the chosen bunch
charge (set of charges from 20 pC to 1 nC with different parameters will be available at the European XFEL).
For example, the pulse duration can be chosen
between few femtoseconds and 100 femtoseconds. In all cases the peak power and the degree of circular polarization will be
comparable to those shown in Table 1. Let us also notice that our method will work in a wide range of
photon energies so that one can easily cover not only L-edges but also M-edges of all interesting elements.
Indeed, in the considered case of lasing at 1.5 nm
the active length of the undulator is 55 m (to be compared to the saturation length of 45 m for the
untapered case, i.e. we have only 20\% increase in length). The total magnetic length of the SASE3 undulator is 105 m so that
there is a big reserve for going to shorter wavelength.
Generally speaking, our method can also work at hard X-ray beamlines if this is requested by users.

Finally, let us note that
in the case of energy loss along the undulator due to the wakefields and spontaneous undulator radiation at high energies,
the strength of the reverse taper can be decreased in accordance with
formula (\ref{beta-gen}), see discussion below. In our case both effects are small corrections, each of them is on the
order of $0.1\%$ in the active part of the SASE3 undulator - to be compared with about $2\%$ of the K change.

\section{A possible operation at LCLS}

A fixed-gap planar undulator
is used to generate hard- and soft- X-ray radiation at the Linac Coherent Light Source (LCLS) \cite{lcls}.
A helical afterburner is going to be installed soon in order to provide a circular polarization for user operation
at LCLS \cite{lcls-polar}.

Design of the planar undulator allows for a mild tapering by making use of canted poles. This option is normally used for compensation
of the beam energy loss along the undulator length, and for the post-saturation taper - in both cases a standard (positive) sign
of taper is needed. We propose here to use a reverse taper to obtain powerful X-ray radiation (in soft- and hard- X-ray regimes)
with a high degree of circular polarization, in excess of 99\%.
Our estimates with the help of the formula (\ref{beta}) suggest that the strength of the reverse taper
should typically be on the order of 1\% over active undulator length.
After optimizing the taper strength and active length of the main undulator,
the K-value of the helical afterburner should be scanned in order to obtain maximum power. Such an experiment can be performed
in the near future.

\section{Some generalizations}

For simplicity we have considered up to now the case when only the undulator parameter K changes linearly along the undulator
length. Obviously, the parameter $\beta$ can be generalized to the case when, in addition, the mean energy of electrons
changes due to the wakefields and spontaneous undulator radiation:

\begin{equation}
\beta =  - \frac{\lambda_{\mathrm{w}}}{4\pi\rho^2} \ \left[ \frac{K(0)}{1+K(0)^2} \ \frac{d K}{d z}
-  \frac{1}{\gamma(0)} \frac{d \gamma}{d z} \right] \ .
\label{beta-gen}
\end{equation}

Here $\gamma(0)$ is the gamma-factor at the undulator entrance. If the energy loss is not negligible, one should decrease
the taper strength correspondingly.

We have mainly considered the case of a SASE FEL in this paper.
In the case of seeded (self-seeded) FELs one can use two modifications of the suppression method: with reverse taper
or with constant detuning of the K parameter (so that the detuning parameter $\hat{C}$ is negative).

We have simulated the helical afterburner for the SASE3 undulator of the European XFEL. Obviously,
as an afterburner one can also consider a cross-planar undulator with a phase shifter \cite{k-j,second-harm}
which may give more possibilities for polarization control. In this case the length of the afterburner should be short enough so that
density modulation stays almost unchanged as the beam propagates in the afterburner. A more complicated cascaded
crossed undulator \cite{tanaka,geloni-pol} can be used as well.

\begin{figure}[tb]

\includegraphics[width=.9\textwidth]{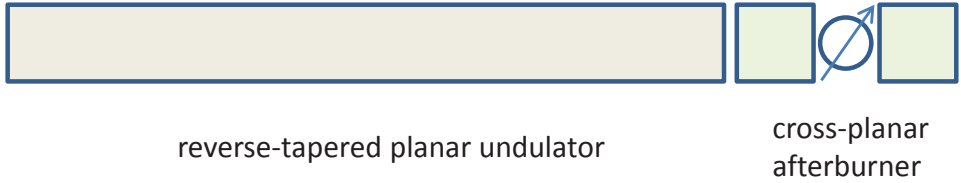}

\caption{\small Scheme for obtaining circular polarization in a cross-planar undulator.}

\label{scheme-cross}
\end{figure}

\section{Acknowledgements}

We would like to thank R.~Brinkmann, V.~Balandin, N.~Golubeva, W.~Decking, and T.~Limberg for useful discussions.

\end{document}